\documentclass[journal=jacsat,manuscript=article]{achemso}

\usepackage[version=3]{mhchem} 
\usepackage[colorlinks=true,urlcolor=black,linkcolor=black,citecolor=black]{hyperref}
\usepackage{ulem}
\usepackage{siunitx}
\usepackage{arydshln}    
\usepackage{bm}
\usepackage{graphicx}
\usepackage{upgreek}
\usepackage{xcolor}
\usepackage{natbib}
\usepackage{array}
\usepackage{float}
\usepackage{booktabs} 
\usepackage[utf8]{inputenc}
\usepackage[T1]{fontenc}

\author{Haiyuan Wang}
\email{haiwa@dtu.dk}
\affiliation[DTU-physics]
{CAMD, Computational Atomic-Scale Materials Design, Department of Physics, Technical University of Denmark, 2800 Kgs. Lyngby, Denmark.}

\author{Nicolas Stenger}
\affiliation[DTU-Electro]
{Department of Electrical and Photonics Engineering
Quantum Photonics of Low-dimensional Systems, Technical University of Denmark, 2800 Kgs. Lyngby, Denmark.}
\alsoaffiliation[NanoPhoton]
{NanoPhoton - Center for Nanophotonics, Technical University of Denmark, Ørsteds Plads 345A, DK-2800 Kgs. Lyngby, Denmark}

\author{Peder Lyngby}
\affiliation[DTU-physics]
{CAMD, Computational Atomic-Scale Materials Design, Department of Physics, Technical University of Denmark, 2800 Kgs. Lyngby, Denmark.}

\author{Mikael Kuisma}
\affiliation[DTU-physics]
{CAMD, Computational Atomic-Scale Materials Design, Department of Physics, Technical University of Denmark, 2800 Kgs. Lyngby, Denmark.}

\author{Jakob Schi\o{}tz}
\affiliation[DTU-physics]
{CAMD, Computational Atomic-Scale Materials Design, Department of Physics, Technical University of Denmark, 2800 Kgs. Lyngby, Denmark.}
\author{Kristian Sommer Thygesen}
\email{thygesen@fysik.dtu.dk}
\affiliation[DTU-physics]
{CAMD, Computational Atomic-Scale Materials Design, Department of Physics, Technical University of Denmark, 2800 Kgs. Lyngby, Denmark.}

\title[An \textsf{achemso} demo]
  {Two-Dimensional Materials as Ideal Substrates for Molecular Quantum Emitters}

\abbreviations{IR,NMR,UV}

\begin{document}


\newpage
\begin{abstract}
The generation and manipulation of non-classical states of light is central to quantum technologies. Color centers in insulators have been extensively studied for single-photon generation, but organic molecules immobilized on substrates have gained attention due to their superior scalability, large oscillator strengths, and tunable emission frequency. Here, we use first-principles calculations to investigate the photoemission from organic molecules adsorbed on 2D materials. Focusing on terrylene on hexagonal boron nitride (hBN), we obtain zero phonon line (ZPL) energies and emission lineshapes in excellent agreement with experiments. Notably, antisite defects in hBN can immobilize the molecule without influencing its key emission features. We further show that the main effect of the 2D substrate is to introduce sharp sidebands near the ZPL as a fingerprint of hindered rotational, translational, or bending modes of the molecule. Our findings provide insight into how substrate interactions shape the optical properties of molecular systems for quantum applications.
\end{abstract}


\section{TOC graph}

\begin{figure}
\centering
\includegraphics[width=8.25cm,height=4.45cm]{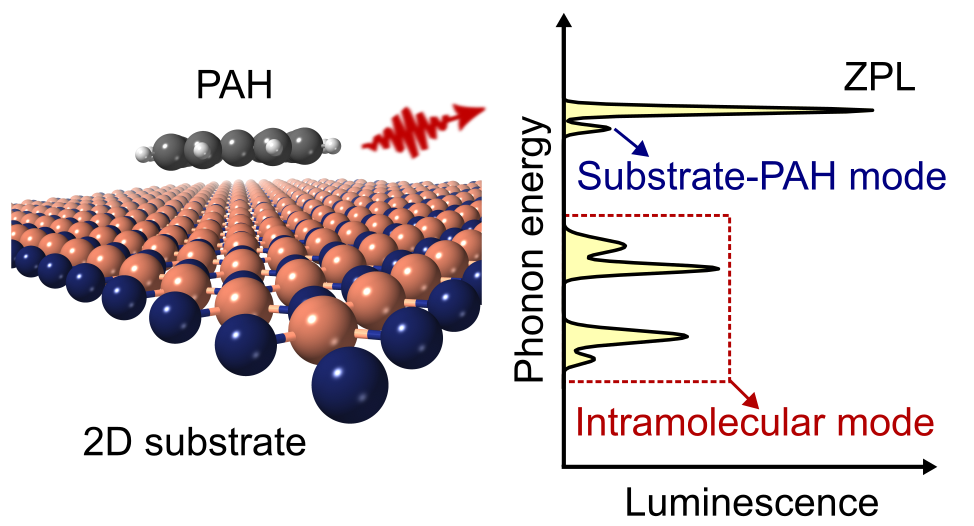}
\end{figure}

\newpage
Single-photon emitters are crucial for several emerging quantum technologies including quantum sensing\cite{degen2017quantum} and quantum communication\cite{gisin2007quantum}. Among the various material platforms currently being explored as the basis for single-photon sources are layered van der Waals (vdW) materials, such as hexagonal boron nitride (hBN)\cite{tran2016quantum} and transition metal dichalcogenides (TMDs)\cite{he2015single, koperski2015single}. These materials can be exfoliated into atomically thin two-dimensional (2D) layers with unique physical properties, which can be enhanced further by combining different 2D layers into vdW heterostructures. Various 2D materials have been found to host quantum emitters in the form of crystal point defects with bright and stable photo-luminescence (PL).\cite{he2015single, tran2016quantum, chakraborty2016localized, sajid2020vncb, fischer2023combining, lee2025room} Recently, Neumann \emph{et al.}\cite{neumann2023organic} proposed that some of the PL spectra observed from hBN crystals could originate from 
polycyclic aromatic hydrocarbon (PAH) molecules, such as terrylene (TRL) and dibenzoterrylene (DBT)~\cite{clear2020phonon}, which are known to form at high temperature in inert atmospheres. 

Single molecules, like PAH, immobilized on a solid-state substrate could offer numerous advantages as single-photon emitters.\cite{toninelli2021single} For example, they are easy to fabricate and control, possess large and well-defined transition dipoles, and can be engineered to emit in specific frequency ranges. Furthermore, single molecules often exhibit sharp zero-phonon lines (ZPLs), leading to exceptionally bright photon sources with high coherence, even at room temperature.\cite{kulzer1999terrylene, lounis2000single, toninelli2021single, schofield2022narrow} On this basis, it appears natural to explore the prospects of single molecule emitters combined with 2D host structures as a platform for single-photon generation. \cite{xu2013graphene, han2021photostable} Recent studies\cite{han2021photostable, smit2023sharp} have measured PL spectra with sharp ZPLs in hBN samples with intentionally embedded TRL molecules. These studies suggest that the interactions between the molecules and the hBN host structure do not influence the molecular emission characteristics much. This situation is radically different from defect centers where the PL lineshape is completely governed by the strong electron-phonon coupling with the host crystal lattice \cite{fischer2023combining,preuss2022resonant}.

To establish a microscopic picture of the photoemission properties of 2D/PAH systems, we use first-principles methods to calculate and analyse the PL spectrum of TRL adsorbed on different 2D materials. The challenge of finding the energetically most favorable adsorption configuration is simplified using the MACE machine learning interatomic potential augmented by the D3 van der Waals (vdW)-correction, in combination with density functional theory (DFT) calculations. For the promising 2D host materials, such as hBN and GaN, we calculate the PL spectrum of adsorbed TRL and compare them to experimental data when available in the literature~\cite{smit2023sharp}. For hBN we furthermore study how point defects in the hBN crystal influence the PL spectrum. This analysis shows that the antisite defects can effectively immobilize the molecule without significantly altering its emission properties. While the ZPL and intramolecular vibronic sidebands are little affected by the 2D crystal, we identify small low-energy sideband(s) with narrow line widths near the ZPL. These sidebands emerge from hindered translational and rotational modes of the molecule when the TRL is adsorbed on pristine hBN, and from out-of-plane bending modes when the TRL adsorbed on defective hBN. This reconfiguration of the low-energy phonon sidebands close to the ZPL could lead to more efficient and coherent single photon sources~\cite{iles2017phonon} for light-based quantum technologies.

\begin{figure}[t]
	\centering
	\includegraphics[width=1\textwidth]{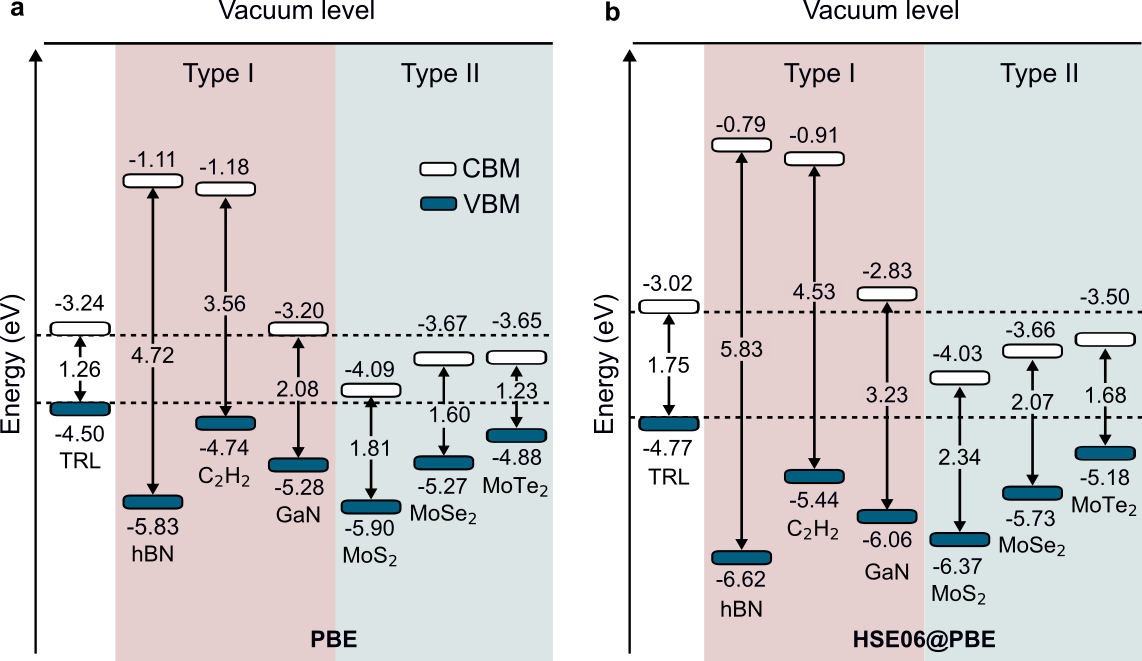}
	\caption{\textbf{Energy level diagram}. Schematic energy-level diagram for the free molecule (TRL) and various 2D materials. The energy levels are calculated with the (a) PBE and (b) HSE06$@$PBE exchange-correlation functionals assuming a common vacuum level. The dash lines represent the HOMO and LUMO energy levels of TRL in the gas phase.}
	\label{figure1}
\end{figure}

A first requirement for an efficient 2D/molecule quantum emitter is that the band gap of the 2D host material straddles the lowest optical transition in the TRL molecule, i.e. the valence band maximum (VBM) and conduction band minimum (CBM) of the host crystal should lie below the highest occupied molecular orbital (HOMO) and above lowest unoccupied molecular orbital (LUMO), respectively. This corresponds to a Type I level alignment, as illustrated in Figure \ref{figure1}. In contrast, for Type II alignment, an excitation of the molecule may decay into a charge-transfer excitation before emission of a photon, which can hinder efficient light emission.

Figure \ref{figure1} presents the level alignment of TRL with six 2D materials, namely hBN, C$_2$H$_2$, GaN, and three Mo-based TMDs. The GaN monolayer is stable in both planar and buckled configurations. \cite{zhuang2013computational, al2016two}, but we focus on the planar configuration. We see that PBE and HSE06@PBE calculations yield consistent results regarding the type of level alignment, which is predicted as Type I for hBN, GaN, and C$_2$H$_2$ and Type II for all the TMDs. Consequently, we focus on the former three interfaces in the following due to their favorable characteristics for efficient light emission. Note that both PBE and HSE06@PBE underestimate the band gap. For example, the experimental electronic band gap of hBN is estimated at 6.25 eV, based on an optical band gap of 5.95 eV and an exciton binding energy of 0.3 eV.\cite{fraunie2025charge} Our HSE06@PBE calculations predict a gap of 5.83 eV, underestimating the experimental value by $\sim$ 0.4 eV. This is consistent with known limitations of hybrid functionals for wide-band-gap materials.
Additionally, the reported electron affinity of hBN varies between positive \cite{marye2024thermal, fraunie2025charge} and negative values \cite{ogawa2019band}, further highlighting the challenges in determining absolute band positions. However, the key focus of our work is on the relative alignment of molecular energy levels with respect to the host materials.

\begin{figure}[t]
	\centering
	\includegraphics[width=1\textwidth]{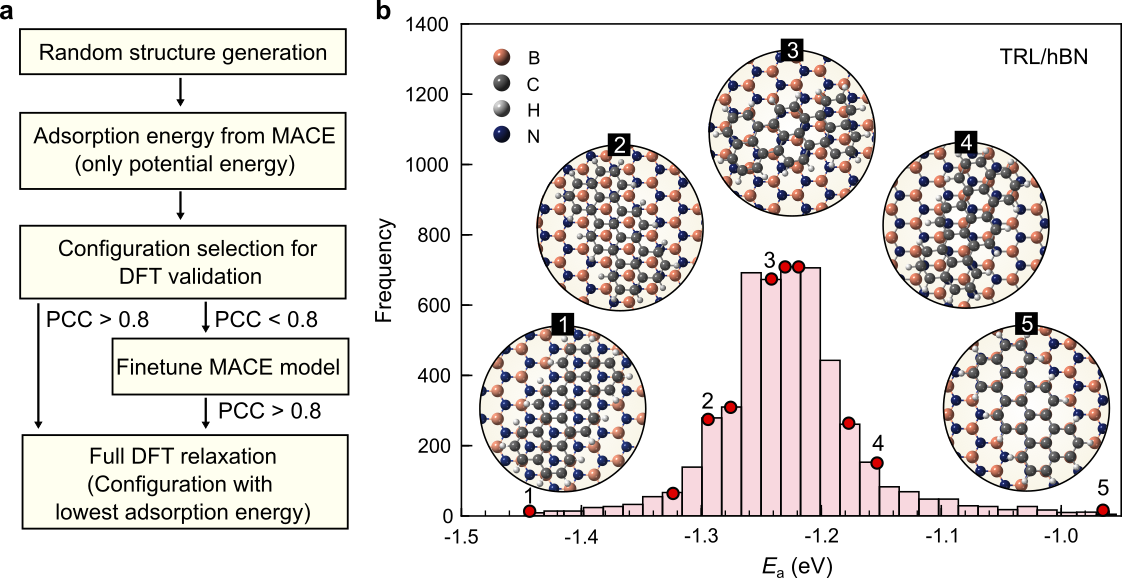}
	\caption{\textbf{Adsorption energies and adsorption configuration}. (a) The workflow followed to determine the most stable adsorption geometry. (b) Histogram showing the adsorption energies calculated using MACE-D3 for the TRL/hBN system. Red circles indicate the structures chosen for further DFT calculations. The atomic structure of five adsorption configurations are also illustrated. The pink, grey, white, and blue spheres represent B, C, H, and N, respectively.}
	\label{figure2}
\end{figure}

To efficiently sample molecular adsorption configuration on 2D substrates, we use the MACE-MP model\cite{batatia2024foundatio}, augmented with a D3 vdW correction (referred to as MACE-D3). For each 2D substrate, we perform the following steps (cf. Figure \ref{figure2}a): (i) Generate a sufficiently dense set of interface configurations (5.000 in this study) by employing a random search strategy where the distance between the molecule and substrate is fixed at 3 Å, while the positions in the \textit{x-y} plane are varied randomly. The rationale for fixing the molecule-substrate distance is to reduce the number of free parameters in the search and focus attention on the variation in adsorption energy caused by the lateral orientation and position of the molecule over the substrate. A distance of 3 Å is chosen as it is a reasonable approximation for the vdW interaction distance.\cite{chen2016capture} (ii) Calculate the single-point total energy for each configuration using MACE-D3, without atomic relaxation. (iii) Select 10 structures spanning the full range of adsorption energies (red points in Figure \ref{figure2}b) for single-point DFT energy calculations, also without ionic relaxation. (iv) Compare the adsorption energies predicted by MACE-D3 with those obtained from DFT. If the Pearson Correlation Coefficient (PCC) is greater than 0.8, the validity of MACE-D3 results is affirmed. If the PCC is less than 0.8, the MACE-D3 model is finetuned on the basis of DFT calculations for randomly selected configurations until the PCC is above 0.8. (v) Perform full DFT relaxation, including force relaxation, on the structure with the lowest adsorption energy predicted by MACE-D3. Our findings suggest that MACE-D3 serves as an efficient tool for single-point calculations, enabling the rapid screening of a large number of configurations (more details shown in the SI).

The optimized TRL/2D interface structures follow particular symmetric patterns despite being governed by non-directional vdW forces. For instance, in hBN and C$_2$H$_2$, the benzene rings of TRL align with those of the substrate, with nitrogen in hBN (or hydrogen in C$_2$H$_2$) positioned at the centers of the TRL hexagons. For GaN and MoS$_2$, the TRL adopts a similar interface configuration, with N in GaN and S in MoS$_2$ relaxing towards the center of the TRL hexagons. Representative structures are shown in Figures 3a,b and S3a,b, respectively.

\begin{figure}[t]
	\centering
	\includegraphics[width=0.95\textwidth]{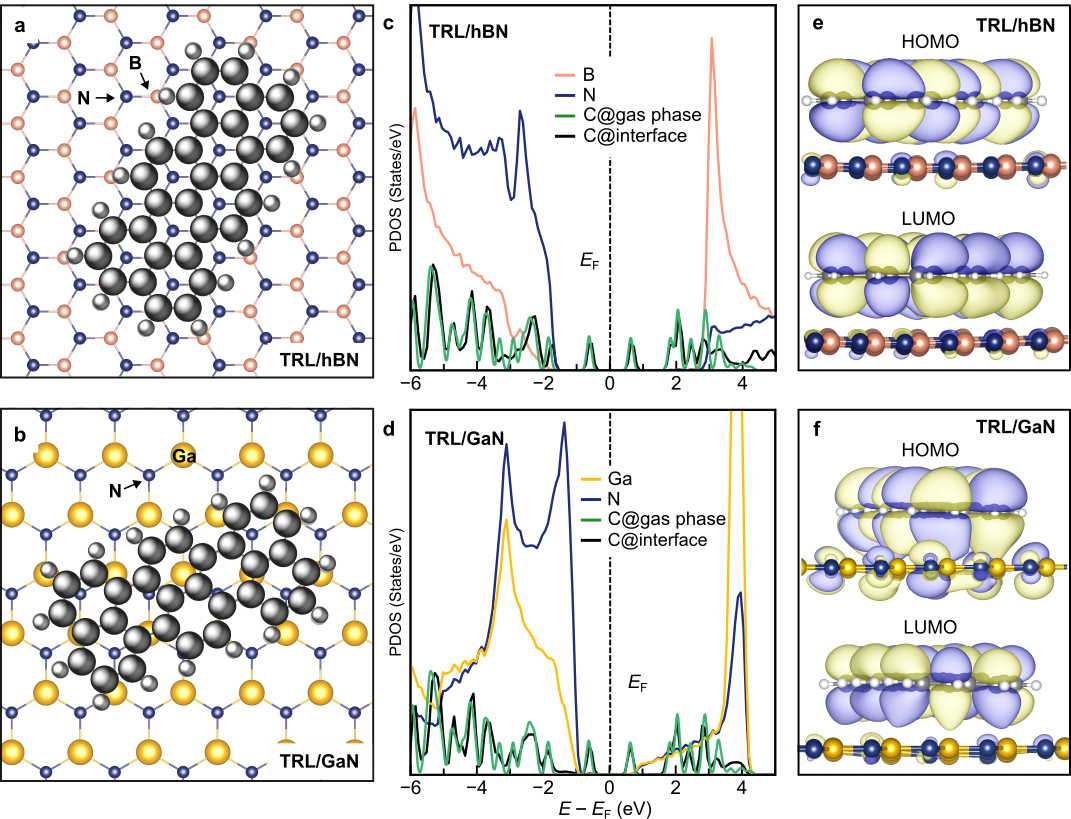}
	\caption{\textbf{Atomic and electronic structures (PBE level)}. Top view of the optimized interface structures for (a) TRL/hBN and (b) TRL/GaN systems. Projected density of states (PDOS) for the (c) TRL/hBN and (d) TRL/GaN systems, with green lines representing the C components from molecules in the gas phase (C@gas phase) for the comparison. The contour plots of the HOMO and LUMO orbitals for (e) TRL/hBN and (f) TRL/GaN systems. The yellow and blue colors represent the positive and negative signs of the molecular orbital wave function.}
	\label{figure3}
\end{figure}

The projected density of states (PDOS) calculated for the optimized TRL/hBN and TRL/GaN structures are shown in Figures \ref{figure3}c,d. Several single-particle eigenstates of the TRL molecule can be seen inside the band gap of the 2D host materials. The energy of these states match very closely with those of the gas-phase molecule suggesting very weak coupling with the 2D states of the substrate (see green curve). In particular, the molecule preserve a closed shell electronic structure with a singlet ground state. Consequently, an accurate determination of the ZPL energy requires spin purification to account for the non-determinantal nature of the HOMO-LUMO singlet transition. The procedure for obtaining the ZPL, including the spin purification corrections with both the PBE and HSE06@PBE functionals, is detailed in the SI, where the HSE06@PBE calculations are carefully benchmarked against results with full HSE06 relaxations for the TRL gas-phase molecule.

\begin{table}[t]
  \caption{\textbf{Key TRL/2D photoluminescence parameters.} Adsorption energy ($E_{\rm a}$), HR factor ($S$), and transition dipole moment ($D$) calculated using the PBE functional. The zero-phonon line ($E_{\rm ZPL}$) energies are determined using both PBE and HSE06@PBE functionals including the spin purification corrections. Radiative rates ($\Gamma$) and lifetimes ($\tau$) are also listed. Corresponding values for the NV$^-$ center, calculated using the same methodology, are included for comparison. Units: All energies are in eV while $D$, $\Gamma$, and $\tau$ are in Debye, $s^{-1}$, and ns, respectively. 
  }
  \resizebox{\textwidth}{!}{
  \label{table1} 
  \begin{tabular}{lcccccclc}
  \toprule
  
   &\multicolumn{4}{c}{PBE}& HSE06@PBE& Expt. \\
   \cline{2-5}  
   & $E_{\rm a}$& $S$& $D$ & $E_{\rm ZPL}$& $E_{\rm ZPL}$& $E_{\rm ZPL}$& $\Gamma$&  $\tau$ \\
    \hline
    TRL gas phase&  $-$& 1.12& 11.2& 1.34& 1.86& --& 1.33$\times$10$^8$& 7.5 \\
    \hline    
     TRL/hBN&       $-$1.87& 1.22& 13.0& 1.34& 1.86& 2.03-2.17$^\textit{a}$& 1.79$\times$10$^8$& 5.6 \\
     TRL/\ce{C2H2}& $-$1.66& 1.16& 10.6& 1.34& 1.86& --& 1.19$\times$10$^8$& 8.4 \\
     TRL/GaN&       $-$1.83& 1.27& 10.5& 1.31& 1.81& --& 1.08$\times$10$^8$& 9.3 \\
    \hline
    TRL/N\textsubscript{B}@hBN& $-$2.49& 1.22& 5.61& 1.32& 1.84& 2.03-2.17$^\textit{a}$& 0.32$\times$10$^8$& 30.9 \\
    \hline
    NV$^-$ center&       & 3.67$^\textit{b}$&& & & &1.25$\times$10$^8$ $^\textit{c}$& 8.0$^\textit{c}$ \\
  \bottomrule  
  \end{tabular} }

  $^\textit{a}$ Ref.~\citenum{smit2023sharp}, 
  $^\textit{b}$ Ref. \citenum{alkauskas2014first}, 
  $^\textit{c}$ Ref.~\citenum{gali2019ab}.
\end{table}

In Table \ref{table1} we report key physical parameters for TRL on three 2D substrates including hBN with a N\textsubscript{B} defect. TRL in the gas phase and NV$^-$ center in diamond are shown for comparison. We first observe that the HSE06@PBE yields a ZPL of 1.86 eV for TRL/hBN. This is in reasonable agreement with the experimental measurements, which range from 2.03 to 2.17 eV \cite{smit2023sharp}. We mark that there is little variation in the ZPL energies between the different 2D substrates. This is partly a result of the weak vdW interactions between the substrate and molecule, which renders orbital hybridisation effects minimal. Even for such weakly bonded interfaces, one could still expect a redshift of the ZPL energy due to image charge effects in the substrate\cite{garcia2011renormalization}. However, because of the similar shapes of the TRL HOMO and LUMO orbitals, the dipole created by the optical transition is very small and the image charge effect becomes negligible. The difference in ZPL energies obtained with PBE and HSE06@PBE is significant. However, the Huang-Rhys (HR) factor, $S$, is accurately captured by PBE by comparison to full HSE06 calculations for the TRL molecule in the gas phase (cf. Table S1). 

Table \ref{table1} shows that all the TRL/2D systems studied exhibit weak electron-phonon coupling as expressed by small HR factors between 1.16 and 1.27. In particular, these $S$ values are significantly lower than that of the prototypical defect center, the NV$^{-}$  in diamond (3.67) \cite{alkauskas2014first}. Furthermore, the calculated transition dipole moments are greater than 10 Debye and are comparable to those of the NV$^-$ center. We note that our calculated lifetimes of 5.6 - 9.3 ns are slightly larger than the experimental value of 3.6 $\pm$ 0.2 ns reported for the TRL/hBN system \cite{smit2023sharp}. Part of this discrepancy can be ascribed to the underestimation of the ZPL energy in our calculations. By adopting the experimental ZPL energy of 2.03 (2.17) eV\cite{smit2023sharp}, we obtain a lifetime of 4.3 (3.5) ns, respectively, in better agreement with the experimental value. Other possible reason for the discrepancy is that our calculations neglect non-radiative decay pathways and temperature effects, which might contribute to a shorter experimental lifetime. 

A recent study by Haas et al. \cite{de2024charge} suggests that TRL diffuses on pristine hBN at elevated temperatures. To explore possible stabilization mechanisms, we consider different point defects in the hBN substrate and assess whether they could immobilize the molecule. Specifically, we consider native antisite defects (N\textsubscript{B} and B\textsubscript{N})\cite{li2025native}, carbon substitutional defects (C\textsubscript{B} and C\textsubscript{N}) \cite{auburger2021towards, jara2021first}, and an oxygen substitutional defect (O\textsubscript{N}) \cite{de2024charge}, which are commonly observed in hBN. Vacancies and O\textsubscript{B} defects are not included, as previous work shows that they do not significantly enhance the adsorption energy \cite{de2024charge}. We find that the MACE-D3 model is unable to describe the local geometric distortions introduced by defects with sufficient accuracy. Consequently, for each defect we use DFT to relax 4-5 configurations with the defect located either underneath or near the molecule. The lowest energy configuration for each defect is illustrated in Figure S6. By analyzing the adsorption energy, ZPL energy, mass-weighted displacement ($\Delta Q$) and $S$ (shown in Table S3), we observe that the N\textsubscript{B} defect decreases the adsorption energy from $-$1.88 eV on the pristine substrate to $-$2.49 eV, thus significantly increasing molecular binding without influencing the electron-phonon interaction, as indicated by the unchanged $S$ of 1.22. Using Arrhenius' law, $\gamma = A e^{-E_{\rm a}/\rm k_{\rm B}T}$, to estimate the rate for jumping from the defect bonding site to a neighboring pristine bonding site we obtain $\gamma=10 \mathrm{s}^{-1}$ at room temperature for an activation energy $E_a = 0.61$ eV and a standard attempt frequency of $A=10^{12} \mathrm{s}^{-1}$. This shows that the molecule can be effectively trapped and that the presence of defects can immobilize or significantly slow the diffusion of molecules. 

We highlight that the optical properties, in particular the ZPL and the HR factor, of the TRL/N\textsubscript{B}@hBN system remain similar to that of TRL on pristine hBN. Although the HOMO of the defective system is delocalized over both the N\textsubscript{B} defect in hBN and the TRL molecule, the LUMO remains entirely localized on the molecule, as shown in Figure S7. The delocalization of the HOMO lowers the transition dipole moment from 13.0 Debye (TRL/hBN) to 5.61 Debye. To understand why the HR factor is almost unaffected by the N\textsubscript{B} defect we decomposed the HR factor for the TRL/N\textsubscript{B}@hBN system into its molecular and substrate parts. This is done by decomposing the displacement vector, $\Delta R$, into its molecular and substrate components and calculate the two corresponding $S$ components. We find that, the molecular component dominates with $S=1.05$, while the substrate contributes only a minor fraction of 0.17. This shows that the reorganization of the structure upon photo-emission mainly involves the atoms of the molecule and explains the 'molecular character' of the optical properties. 
From atomic perspective, Figure S6 shows the N\textsubscript{B} defect does not alter the geometry of the molecule, unlike B\textsubscript{N} defects, which causes noticeable structural distortion and explains the increase in $\Delta Q$ and $S$. Such defect-induced changes in optical properties are common. As shown in Table S3, the ZPL, $\Delta Q$ and $S$ vary significantly for TRL on hBN substrate with different defect types. While we do not propose that the specific N\textsubscript{B} defect necessarily plays a role in the experimental work\cite{smit2023sharp}, our calculations show that it is possible to immobilize an organic molecule via crystal defects in the hBN without significantly altering the interactions between the phonons and molecular transitions.

\begin{figure}[t]
	\centering
	\includegraphics[width=0.88\textwidth]{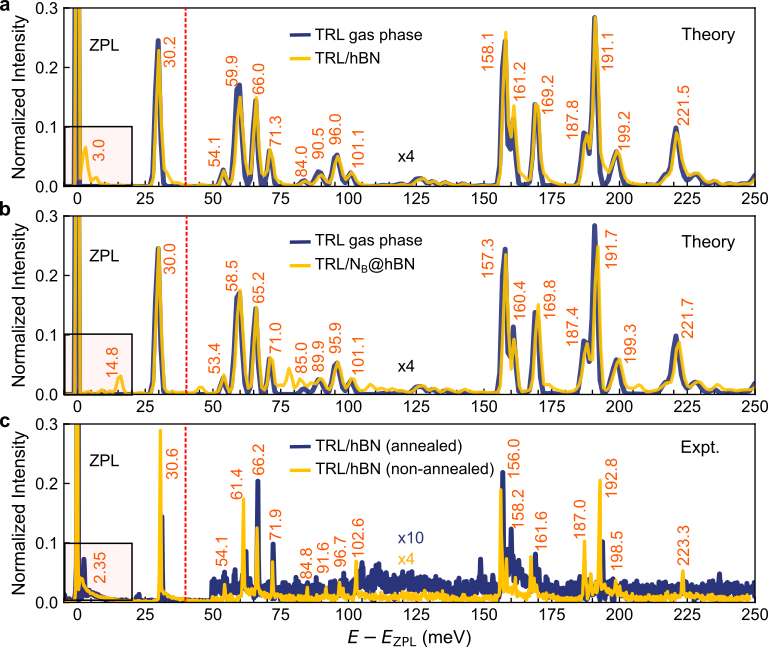}
	\caption{\textbf{Photoluminescence spectra.} {Calculated photoluminescence line shapes for an isolated TRL molecule, TRL adsorbed on pristine hBN, and TRL on a nitrogen antisite defect in hBN. The lower panels shows experimental spectra from Ref. \citenum{smit2023sharp} of TRL on annealed and non-annealed hBN for comparison. All spectra are normalized such that the intensity of the ZPL is set to unity, with the ZPL positions shifted to 0. The spectral region beyond 45 meV has been magnified by the factor indicated in the figure for better visibility.}}
	\label{figure4}
\end{figure}

\begin{figure}[t]
	\centering
	\includegraphics[width=0.98\textwidth]{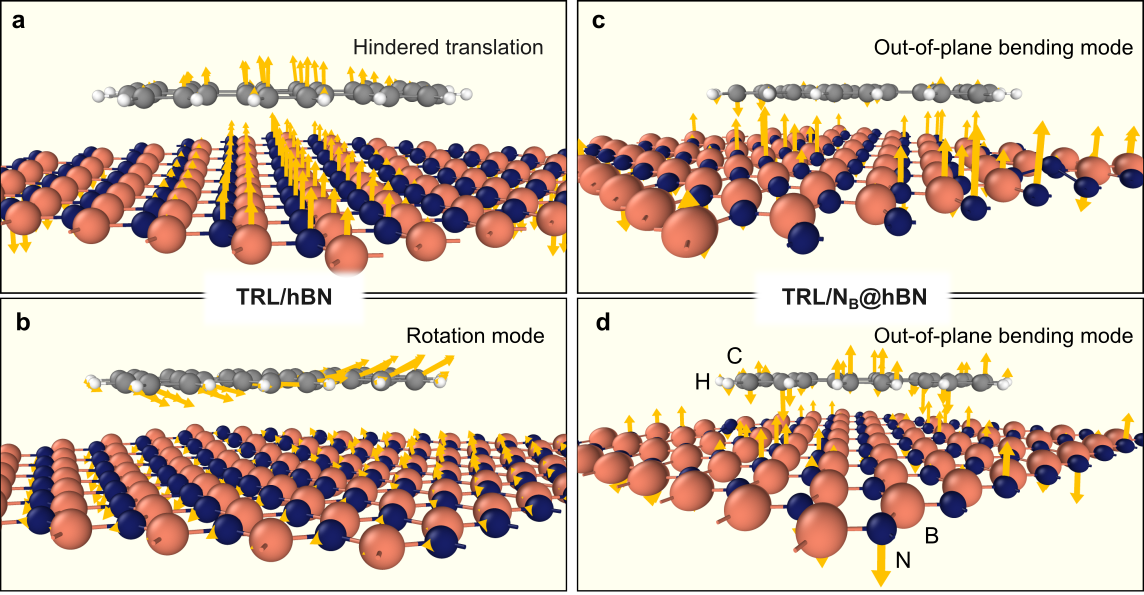}
	\caption{\textbf{Vibrational modes} {The atomic displacement vectors of the two vibrational modes with the lowest frequency in Figure \ref{figure4}(a) and (b) (red region) for TRL/hBN and TRL/N\textsubscript{B}@hBN are as follows: For TRL/hBN, the two modes correspond to the molecule performing (a) a hindered out-of-plane translation ('breathing mode') and (b) a hindered out-of-plane rotation ('flipping mode'). For TRL/N\textsubscript{B}@hBN, the modes correspond to the out-of-plane bending motions.}}
	\label{figure5}
\end{figure}

PL spectra are calculated from the phonon spectral densities using the generating function method \cite{alkauskas2014first, tawfik2022pyphotonics}. Figure \ref{figure4} presents the normalized PL spectra for isolated TRL, TRL/hBN, and TRL/N\textsubscript{B}@hBN. The experimental spectra for a molecule adsorbed on non-annealed hBN and hBN annealed in an oxidizing atmosphere\cite{smit2023sharp} are also shown for comparison. We observe an excellent agreement between the calculated and experimental data. In particular, both position and relative magnitude of all significant side bands are well reproduced. The side bands above 25 meV are almost identical for the isolated and adsorbed molecule, indicating that they originate from intramolecular vibrations. The same trend is seen for TRL on pristine GaN and on pristine C$_2$H$_2$ (see Figure S9). The main qualitative difference between the isolated and adsorbed molecule occurs in the low-energy side bands near the ZPL (highlighted in the red box). Both the calculated and experimental spectra of TRL/hBN show a distinct peak around 2-3 meV above the ZPL, which is absent in the theoretical spectrum of the isolated molecule. We ascribe the emergence of this sharp low energy side band peak to hindered translation and rotation modes of the adsorbed molecule for the TRL/hBN case and to out-of-plane bending modes for the TRL/N\textsubscript{B}@hBN (see Figure \ref{figure5}). We notice that the low-energy peaks highlighted by red boxes are absent in the calculated spectra reported by Haas et al.\cite{de2024charge} We hypothesize that this discrepancy may arise from the use of the ONIOM method in their study, where the hBN substrate is treated using the semi-empirical PM3 model. As a result, the interactions between TRL and hBN may not have been fully captured. In contrast, our calculations treat both molecule and the substrate at the same \textit{ab inito} level.

Furthermore, the experimental PL spectrum of TRL on non-annealed hBN in Figure \ref{figure4}c shows a low-energy shoulder extending to around 10 meV from the ZPL; this shoulder is missed by the DFT calculations. We speculate that the shoulder is due to low-energy hindered rotational/translational modes of loosely bound molecules on many different adsorption sites. This interpretation is also consistent with the fact that the shoulder is suppressed on the annealed and thus cleaner hBN, and instead is replaced by a more well defined peak presumably originating from molecular vibrations across more well defined and similar adsorption configurations. In general the conversion of a broadband, continuous shoulder into a sharp distinct peak is advantageous for single-photon generation as it allows for more well defined frequency filtering of the photons emitted in the ZPL leading to improved quantum coherence such as indistinguishability~\cite{iles2017phonon}.   

The phonon spectral functions are shown in Figure S10. Their peaks closely resemble those of the PL spectrum indicating that the photon emission mainly involve excitation of single phonons. As compared to TRL on hBN, the GaN substrate produces more coupled substrate-molecule modes that contribute to the peak near the ZPL in the range of 2-15 meV, while the hindered translation and rotation modes still remain dominant. In contrast, TRL/C$_2$H$_2$ exhibits only one hindered rotational mode of the molecule, which is responsible for the small peak near the ZPL. For all 2D materials studied in this work, we observe a reorganization of the phonon side band close to the ZPL. This result indicates that the broadband low-energy phonon side bands, usually detrimental to the quantum coherent properties of single photons, could be mitigated via molecule-substrate interaction-engineering.

In summary, we have used \textit{ab inito} calculations to establish a microscopic picture of photoemission from the polycyclic aromatic hydrocarbon (PAH) molecule terrylene (TRL) adsorbed on different pristine and defective 2D materials. The calculated photoluminescence spectra for TRL in the gas phase and adsorbed on hexagonal boron nitride (hBN) are found to be in excellent agreement with experiments. Our calculations further show that point defects in hBN can immobilize TRL at room temperature with negligible effect on the PL lineshape. Analysis of the phonon side bands in the PL spectrum shows that the main effect of the 2D substrate is to suppress the low-energy shoulder present in molecules adsorbed on unclean substrates and instead introduce a discrete peak originating from hindered rotation and translation modes of the immobilized molecule. This sharp resonance close to the ZPL can be easily filtered out in order to improve coherence of the single photons for future applications in light-based quantum technologies. These findings contribute to a broader understanding of the interplay between PAH molecules, 2D substrates, and defects. This can open up new directions for future researches, such as screening molecule databases, which will accelerate the discovery of single-photon emitters.

\textbf{Computational Details} All the calculations are performed using Vienna \textit{ab initio} simulation package (VASP)\cite{kresse1994ab,kresse1996efficient} based on density functional theory (DFT) \cite{hohenberg1964inhomogeneous,kohn1965self} with the projector augmented wave (PAW) method.\cite{blochl1994projector} Constrained DFT \cite{kaduk2012constrained,gali2009theory} is used for the excited state calculations. This method is also referred to as $\Delta$SCF (Delta Self-Consistent Field) \cite{kaduk2012constrained,jones1989density} when calculating the total energy difference between the ground and excited state. All the calculations are spin-polarized, and the geometries are fully relaxed with van der Waals correction (vdw-D3).\cite{grimme2011effect,grimme2010consistent} Dipole corrections are switched on for slabs to avoid the spurious interactions between periodic images.\cite{neugebauer1992adsorbate} The semi-local exchange-correlation functional of Perdew, Burke, and Erzenerhof (PBE) \cite{perdew1996rationale} is employed to optimize the geometries. The total energy convergence and structure minimization criterion are set at $10^{-7}$ eV and $10^{-3}$ eV\text{\AA}$^{-1}$, respectively. For the ZPL, we also use HSE06@PBE functional. In this approach, the electronic properties are established through self-consistent one-shot calculations employing the standard Heyd, Scuseria, and Ernzerhof (HSE06) \cite{krukau2006influence}, with $\alpha=0.25$, and optimized geometries using PBE. We employ 10$\times$10$\times$1 supercells for hBN and C$_2$H$_2$, and 8$\times$8$\times$1 for TMDCs and GaN, ensuring that the minimum distance between molecules exceeds 10 \text{\AA}. The plane-wave cutoff energy of 520 eV and a gamma k-point grid are adopted to guarantee the accuracy of the results. The gas phase molecules are simulated by confining them within a box. The phonons of the supercell are calculated using the finite displacement method under the PBE functional. We assume that the phonons remain the same in the ground and excited states. This assumption forms the basis of the method used to calculate the PL spectrum. The host 2D materials are obtained from Computational 2D Materials Database (C2DB): https://c2db.fysik.dtu.dk/.\cite{haastrup2018computational} 

The MACE-MP model used is the "large" model \cite{batatia2024foundatio}, which employs message parsing with an equivariance degree of order L=2. To ensure consistency with the VASP setup, we apply the zero-damping D3 dispersion correction in the MACE calculations, using cutoff radius of 50.2 Å for pair interactions and 20 Å for coordination numbers. Before finetuning, the D3 dispersion contribution was subtracted from the 60 single-point DFT calculations, and the MACE-MP model was finetuned for an additional 20 epochs. All other hyperparameters remained identical to those used in the initial training of the “large” MACE-MP model\cite{batatia2024foundatio}. The D3 dispersion contribution was reintroduced when applying the finetuned model.

\begin{acknowledgement}
The authors thank T. Boland, and Y. W. Li for their useful discussions. K.S.T. acknowledges support from the Novo Nordisk Foundation Challenge Programme 2021: Smart nanomaterials for applications in life-science, BIOMAG Grant No. NNF21OC0066526, and the Novo Nordisk Foundation Data Science Research Infrastructure 2022 Grant:  A high-performance computing infrastructure for data-driven research on sustainable energy materials, Grant no. NNF22OC0078009. K.S.T. is a Villum Investigator supported by VILLUM FONDEN (grant no. 37789). N.S. is supported by the Danish National Research Foundation through NanoPhoton - Center for Nanophotonics, Grant No. DNRF147 and by the Novo Nordisk Foundation NERD Programme (project QuDec NNF23OC0082957). 
\end{acknowledgement}

\begin{suppinfo}

The Supporting Information is available free of charge at https://pubs.acs.org.\\
Adsorption configuration determination; Spin purification; Functional determination; TRL adsorbed on defective hBN; Emission spectra for TRL/GaN and TRL/C$_2$H$_2$; Spectral function; Determination of key parameters

\end{suppinfo}

\bibliography{achemso-demo}

\end{document}